\documentclass[12pt,superscriptaddress]{iopart}

\usepackage{fancyhdr} 
\newtheorem{thm}{Theorem}%[section]

\newtheorem{definition}{Definition}

\usepackage{yfonts}

\fancypagestyle{plain}{% 
\fancyhead{} %get rid of the headers on plain pages 
}

\usepackage{iopams}
\usepackage{bm}
\usepackage{graphicx}
\usepackage{showlabels}
\usepackage{amssymb}
\usepackage{hyperref}

\def\la{\label}
\def\be{\begin{equation}}
\def\beq{\begin{equation}}
\def\eeq{\end{equation}}
\def\ee{\end{equation}}
\def\bea{\begin{eqnarray}}
\def\eea{\end{eqnarray}}
\def\p{\partial}

\begin{document}

\title{The Lagrangian and symplectic structures of the Kuramoto oscillator model}

\author{Sherwin Kouchekian and Razvan Teodorescu}
\address{Department of Mathematics and Statistics, University of South Florida, Tampa, FL 33620}
\eads{\mailto{skouchekian@usf.edu}, \mailto{razvan@usf.edu}}

\begin{abstract}
Despite being under intense scrutiny for 50 years, the Kuramoto oscillator model has remained a quintessential representative of non-equilibrium phase transitions. One of the reasons for its enduring relevance is the apparent lack of an optimization formulation, due to the fact that (superficially), the equations of motion seem to not be compatible with a Lagrangian structure. We show that, as a mean-field classical (twisted) spin model on $S^2$, the Kuramoto model can be described variationaly. Based on this result perturbation analysis around (unstable) Kuramoto equilibria are shown to be equivalent to low-energy fluctuations of mean-field Heisenberg spin models. Intriguingly, off-plane perturbations around these equilibria configurations turn out to be described 
%In this form, the model is shown to enjoy 
by a semiclassical Gaudin model, pointing to the fact  that oscillator synchronization maps to the spin pairing mechanism investigated by Richardson and subsequently by others.   
\end{abstract}

\pacs{05.30, 05.40, 05.45}
\submitto{\JPA}

\maketitle

%\newpage

%\tableofcontents

%\newpage

\section{Introduction} \label{sec:intro}

Since its introduction  50 years ago, the Kuramoto oscillator model has been considered a fundamental example for dynamical systems exhibiting non-equilibrium phase transitions, in this case the synchronization phase transition. After reviewing the main characteristics of this system, we show that, in the original angular variables, the Langevin-type Kuramoto equations are not compatible with a Lagrangian structure, which is a main reason the problem was not given a variational formulation throughout the years. 

We then explore a complex bilinear structure which allows to retrieve most of the known exact results about this model. While still not Lagrangian, this exploration leads to the formulation of the problem as the mean-field limit of a spin system in $\mathbb{R}^3$, whose two-dimensional restriction reduces to the Kuramoto model. We therefore derive its corresponding Lagrangian and Hamiltonian, which allows to explore the planar and off-planar perturbations around Kuramoto equilibria. The planar perturbations confirm the known stability analysis of synchronization states, as a direct corollary of the exact Hamiltonian for the generalized model. The off-plane perturbations turn out to be described by a semiclassical Gaudin model with energy spectrum given by the centered frequency distribution of the Kuramoto oscillators. This result establishes a natural relationship between Kuramoto oscillator synchronization, on one hand, and (Anderson) pseudo-spin pairing in the semiclassical Gaudin model, on the other. 

\section{A brief review of the Kuramoto model} \la{first}

\subsection{Oscillators interacting via a mean-field attractive force - a first analysis of the phase-locking and synchronization phenomena.}

Let $\theta_j(t), \, j = 1, 2, \ldots, N$ represent the angular variables of $N$ oscillators, with (arbitrary) corresponding natural frequencies $\omega_j \in \mathbb{R}_+$. Denoting by $\lambda \ge 0$ the {\it{coupling constant}} of the mean-field attractive interaction between the oscillators, the Kuramoto equations take the Langevin form

\be \la{eq_1}
\dot{\theta_j} = \omega_j + \frac{\lambda}{N}\sum_{k=1}^N \sin(\theta_k - \theta_j) := F_j(\{ \theta_k \}_{k=1}^N), \,\,  j = 1, 2, \ldots, N,
\ee

\noindent and where $\dot{f}$ represents the time derivative of the function $f(t)$. 

Two oscillators, identified by their indices, $\{ j, \ell \}$, are said to be {\it{phase-locked}} if their angular difference is a constant: 

\be \la{eq_2}
\dot{\theta_j} - \dot{\theta_\ell} = 0 \Leftrightarrow \omega_j + \frac{\lambda}{N}\sum_{k=1}^N \sin(\theta_k - \theta_j) = 
\omega_\ell + \frac{\lambda}{N}\sum_{k=1}^N \sin(\theta_k - \theta_\ell)
\ee

\noindent If oscillators $j, \ell$ are phase-locked, their phase difference, denoted by $\Delta_{j\ell} := \theta_j - \theta_{\ell} = -\Delta_{\ell j}$, will be constant. This difference can be identified to the phase difference between the two angles, stationary  in a co-rotating frame turning at an angular speed equal to the common value $\dot{\theta_j} = \dot{\theta_\ell}$. It is then an obvious proof by contradiction that, for phase-locked oscillators of different natural frequencies, $\omega_j \ne \omega_\ell$, their two phases cannot coincide: 

\be \la{eq_3}
\dot{\theta_j} = \dot{\theta_\ell}, \,\, \omega_j \ne \omega_\ell \Rightarrow \Delta_{j\ell} \ne 0. 
\ee

Let us notice that a trivial estimate performed in Eq.~\ref{eq_2} yields the following necessary condition for the phase locking of two oscillators, $j, \ell$, when $N \ge 3$:

\be \la{eq_4}
\omega_j - \omega_{\ell} = \lambda \left [ 
\frac{\sum_{k=1}^N \sin(\theta_k - \theta_\ell) - \sin(\theta_k - \theta_j)}{N}
\right ] \Rightarrow \lambda \ge \frac{N| \omega_j - \omega_{\ell} |}{2(N-1)}
\ee

We conclude that, for a system of $N \ge 3$ oscillators in which all the frequencies are distinct, for values of the coupling constant below a {\it{critical value}} $\lambda_c$

$$
\lambda < \lambda_c : =  \frac{N}{2(N-1)} \min_{1 \le j\ne \ell \le N}| \omega_j - \omega_\ell |,
$$ 

\noindent the system will always remain unlocked, while for values of the coupling constant exceeding a {\it{synchronization value}} $\lambda_s$

$$
\lambda \ge \lambda_s :=   \frac{N}{2(N-1)}  \max_{1 \le j\ne \ell \le N}| \omega_j - \omega_\ell |,
$$

\noindent  the system may become completely phase locked, for special configurations of the relative phases, $\{ \Delta_{j \ell} \}_{j, \ell = 1}^N$. In particular, a Kuramoto system in which oscillators have the same, common frequency $\omega_ j = \Omega, \, \forall \, j = 1, 2, \ldots, N$ (for which $\lambda_c = 0 = \lambda_s$) will always possibly fully phase-lock at arbitrarily small values of the coupling constant, $\lambda > 0$.  

In the special case $N=2$, we obtain the reduction 

$$
\lambda_c = \lambda_s = |\omega_1 - \omega_2|, 
$$

\noindent with the phase-locking condition 

$$
\sin \Delta_{12} = \frac{\omega_1 - \omega_2}{\lambda}.
$$

In the general case $N \ge 3, \, 0 < \lambda_c < \lambda < \lambda_s$, the system will exhibit a mixture of the two phases, with a relative proportion of phase-locked oscillators depending on the precise distribution of the frequencies, relative to the values $\lambda, \lambda_c, \lambda_s$. 

These features (the system exhibiting a mixture of phases while in a dynamic, time-dependent state, and possibly also depending on initial conditions through the relative phases $\Delta_{j \ell}$) have since been generalized as generic characteristics of non-equilibrium phase transitions. It should be emphasized that this class of phenomena, in contrast with the much better-established theory of equilibrium phase transitions, in which the existence of a well-defined globally convex functional (a thermodynamic potential) and its degree of smoothness allow to classify and indeed fully describe the critical properties of a system (such as in first-order phase transitions, second-order, etc.)  fundamentally lacks a variational formulation, as we demonstrate in the next section.

\subsection{Analyzing variational presentations of the Kuramoto model in two dimensions}

\subsubsection{Non-Lagrangian nature of the original Kuramoto equations.}

Let us first note that, superficially, Eqs.~\ref{eq_1} cannot be realized as Euler-Lagrange equations for a Lagrangian density ${L}(\theta_j, \dot{\theta}_j)$: taking one derivative in Eqs.~\ref{eq_1}, we obtain

\be \la{eq_5}
\ddot{\theta_j} = %\sum_{k=1}^N \dot{\theta_k} \frac{\p F_j}{\p \theta_k} = 
\frac{\lambda}{N}\sum_{k=1}^N \cos(\Delta_{kj}) \left [
\omega_k - \omega_j + \frac{\lambda}{N}\sum_{\ell = 1}^N \sin(\Delta_{\ell k}) - \sin(\Delta_{\ell j}) 
\right ]
\ee 

\noindent Assuming that Eqs.~\ref{eq_5} are Euler-Lagrange equations for  ${L}(\theta_j, \dot{\theta}_j)$, it would follow that 

\be \la{eq_6}
 {L}(\theta_j, \dot{\theta}_j) = \sum_{j=1}^N %\left [ 
 \frac{1}{2}\dot{\theta}^2_j  
 %\right ]
 + \Phi(\{\theta_k\}_{k=1}^N), 
\ee

\noindent since the Lagrangian cannot contain  terms of the form $\dot{\theta_j^2}$ (which would generate a first-order derivative in Eqs.~\ref{eq_5}), as they would only have a boundary contribution to the action, therefore no effect under the Euler-Lagrange first variation. We are therefore seeking a function $ \Phi(\{\theta_k\}_{k=1}^N)$ such that

\be \la{eq_7}
\frac{\p \Phi}{\p \theta_j} = 
\frac{\lambda}{N}\sum_{k=1}^N \cos(\Delta_{kj}) \left [
\omega_k - \omega_j + \frac{\lambda}{N}\sum_{\ell = 1}^N \sin(\Delta_{\ell k}) - \sin(\Delta_{\ell j}) 
\right ]
\ee

We prove that this is inconsistent by verifying that, for general $j \ne q$, 

\be \la{eq_8}
\frac{\p^2 \Phi}{\p \theta_j\p \theta_q} \ne 
\frac{\p^2 \Phi}{\p \theta_q\p \theta_j}.
\ee

From Eq.~\ref{eq_7}, we have %(since $j \ne q$)

$$
\frac{\p^2 \Phi}{\p \theta_q\p \theta_j}  = 
\frac{\lambda}{N}(\omega_j - \omega_q) \sin(\Delta_{qj})  + 
\left (\frac{\lambda}{N}\right )^2
\sum_{k, \ell = 1}^N  \sin(\Delta_{jk}) \left [\sin(\Delta_{lk}) - \sin(\Delta_{\ell j}) \right ]
 [\delta_{k q} - \delta_{jq} ]
$$

$$
+ \left ( \frac{\lambda}{N}\right )^2
\left \{
\sum_{k, \ell =1}^N \cos(\Delta_{kj})
\left [
\cos(\Delta_{\ell k}) (\delta_{\ell q} - \delta_{kq})  
- \cos(\Delta_{\ell j}) (\delta_{\ell q} - \delta_{jq}) 
\right ]
\right \}
$$

Therefore, for $j \ne q$, the difference 

$$
\frac{\p^2 \Phi}{\p \theta_j\p \theta_q} - \frac{\p^2 \Phi}{\p \theta_q\p \theta_j} = 
\left ( \frac{\lambda}{N}\right )^2
 \cos(\Delta_{qj}) \sum_{k=1}^N 
 \left [
  \cos(\Delta_{k j}) - \cos ( \Delta_{k q}) 
 \right ], 
% \left ( \frac{\lambda}{N}\right )^2 \cos(\Delta_{qj})  \sum_{l=1}^N [\cos(\Delta_{lq}) - \cos(\Delta_{lj})]
$$

\noindent which cannot be 0 unless for very special values of $\{ \theta_j \}, j \in \Sigma_N$. We conclude that, in its original angular variables,   the close-form conditions do not hold (Eqs.~\ref{eq_8}), and the Kuramoto equations Eq.~\ref{eq_5} are not of Euler-Lagrange form. 

\subsubsection{A first Poisson bracket structure for Kuramoto oscillators.}

Introducing the notation $z_j(t) := e^{i \theta_j(t)}, \,\, j \in \Sigma_N := \{ 1, 2, \ldots, N \}$, we arrive at the following formulation for the Kuramoto equations:

\be \la{eq_9}
\dot{z}_j = z_j \left [ i\omega_j + \frac{\lambda}{2N}\sum_{k=1}^N (z_k \bar{z}_j - \bar{z}_k z_j) \right ]
\ee

\noindent Defining the following  antisymmetric and  symmetric  bilinear forms on $\mathbb{C} \times \mathbb{C} \to \mathbb{R}$, 

\be \la{eq_10}
[u, v] := \frac{1}{2i} (u \bar v - v \bar u), \quad \{ u, v \} := \frac{1}{2} (u \bar v + v \bar u),  
\ee

\noindent as well as the {\it{complex order parameter}}, 

$$
r(t) = \frac{1}{N}\sum_{k=1}^N z_k(t), 
$$

\noindent we arrive at the Kuramoto equations in complex form: 

\be \la{eq_11}
\dot{z}_j = i z_j (\omega_j + \lambda [r, z_j] )
\ee

We will denote the modulus and argument of the complex order parameter as
$$
r(t) = |r|(t) e^{i \theta_0(t)} \Rightarrow 0 \le |r|(t) \le 1, 
$$
\noindent and locking of oscillators with the order parameter variable is defined the same way as already introduced. 

\begin{thm} \la{Thm_properties}
Let $\alpha \in \mathbb{R}$ and $u, v \in \mathbb{C}^*$. We list the following properties of the bilinear forms defined in Eq.~\ref{eq_10}:

\be \la{Lie}
[u, v] = -[v, u], \quad [u, [v, w]] + [w, [u, v]] + [v, [w, u]] = 0,
\ee
%\be \la{eq_12}
\be \la{brackets}
\{u, v\} = \{v, u\}, \quad
[\alpha u, v] = \alpha [u, v], \quad 
%[u, \alpha v ] = {\alpha} [u, v], \quad 
[i \alpha u, v] = i\alpha  \{ u, v \}, 
\ee
%\ee
\be \la{eq_zeros}
[u, v] = 0 \Leftrightarrow \arg \frac{u}{v} \in \{ 0, \pi \}, \quad \{u, v\} = 0 \Leftrightarrow \arg \frac{u}{v} \in \left \{-\frac{\pi}{2}, \frac{\pi}{2}  \right \}.
\ee
\end{thm}

\noindent The proof is based on elementary computations. This formalism allows to retrieve in a convenient manner conditions for phase-locking of two or several oscillators: 

\begin{thm} \la{Phase-locking conditions}
Assume that oscillators $\{j, k\} \subseteq \Sigma_N$ are phase-locked, then
$$
\left [z_j - z_k, r \right ] =  \frac{\omega_j - \omega_k}{\lambda} 
$$
Furthermore, if the two locked oscillators have the same natural frequency, $\omega_j = \omega_k$, this implies $\frac{z_j - z_k}{r} \in \mathbb{R}$, thus $z_j, z_k,$ and $r$ are all phase-locked.
\end{thm}

%\noindent Taking one derivative, we obtain
%
%$$
% \left [\frac{\dot z_j - \dot z_k}{\omega_j - \omega_k}, r \right ] = 0 \Rightarrow 
% \frac{1}{\omega_j - \omega_k} \left (
% \{ \omega_j z_j, r \}  -  \{ \omega_k z_k, r \} + 
% \lambda \{ [r, z_j], r \} -  \lambda \{ [r, z_k], r \}
%\right )
%= 0
%\{ \omega_j z_j -   \omega_k z_k + \lambda  [r, z_j-z_k], r \} = 
%0
%$$

\noindent {\it{Proof. }}
From Eq.~\ref{eq_11}, we find that 
$$
\frac{\dot z_j }{z_j} = \frac{\dot z_k}{z_k} \Rightarrow \left [\frac{z_j - z_k}{\omega_j - \omega_k}, r \right ] =  \frac{1}{\lambda} 
$$

\noindent Using the homogeneity properties of the antisymmetric form, we find the result. If the natural frequencies are also equal, this becomes

\be \la{eq_two}
\left [{z_j - z_k}, r \right ] = 0 \Rightarrow \frac{z_j - z_k}{r} \in \mathbb{R}, 
\ee

\noindent and by Eq.~\ref{eq_zeros}, the phases of $z_j, z_k,$ and $r$ are all locked. If we denote by $\varphi_j$ the relative phase difference of $z_j$ with respect to $r$, then $\varphi_k$ can only take the values $\varphi_j$ or $\pi - \varphi_j$. 

Evidently, this generalizes to all the oscillators, so that all phase-locked oscillators sharing the same natural frequency coalesce at the same one point $\frac{zr}{|r|} \in S^1$ or at two points $\frac{zr}{|r|}, -\frac{\bar{z}r}{|r|} \in S^1$, fixed relative to the direction of the variable $r$.   \hfill $\square$

%
%Also, for $N=2$, phase-locking of  the oscillators of frequencies $\omega_1 \ne \omega_2$ implies 
%
%%$$
%\be \la{eq_N_2}
%\left [\frac{z_1 - z_2}{\omega_1 - \omega_2}, \frac{z_1 + z_2}{2} \right ] =  \frac{1}{\lambda} \Rightarrow 
%[z_1, z_2] = \frac{\omega_1 - \omega_2}{\lambda} \Rightarrow 
%\sin \Delta_{12} = %\left ( 
%\frac{\omega_1 - \omega_2}{\lambda}  
%%\right )
%\ee
%%$$
%
%[TO BE EXPANDED] Notice that at the onset of phase-locking, for $\lambda = \lambda_c$, $\Delta_{12} = \pm \frac{\pi}{2}$. As $\lambda$ is increased from the critical value, if this describes the onset of synchronization (which would require the presence of other oscillators, such that $|r| = 0$ at $\lambda = \lambda_c$), then the local expansion will lead to the known mean-field scaling law 
%
%$$
%|r|^2 \sim \frac{\lambda - \lambda_c}{\lambda_c}
%$$

\noindent 
Note that, if three oscillators of distinct frequencies $\omega_j, \omega_k, \omega_\ell$ phase-lock, Eq.~\ref{eq_two}  imply
$$
\left [\frac{z_j - z_k}{\omega_j - \omega_k} - \frac{z_k - z_\ell}{\omega_k - \omega_\ell} , r \right ] = 0
\Rightarrow
\frac{(\omega_k - \omega_\ell)z_j + (\omega_\ell - \omega_j)z_k +  (\omega_j - \omega_k)z_\ell}{r} \in \mathbb{R}.
$$ 
\noindent Equivalently, using the angle differences $\varphi_j = \arg(z_j) - \arg(r)$, 
%$$

\be \la{eq_three}
(\omega_k - \omega_\ell)\sin \varphi_j + (\omega_\ell - \omega_j)\sin \varphi_k +  (\omega_j - \omega_k)\sin \varphi_\ell = 0,
\ee
%$$  
\noindent which admits the class of solutions of the form
\be \la{sol_3}
\sin \varphi_j = \alpha(\omega_j - \beta),
\ee
\noindent with $\alpha, \beta$ real constants chosen for proper normalization. 

\subsubsection{The special case of identical oscillators.}
 
These results generalize to finding the necessary condition for a system of oscillators of identical frequencies to phase-lock:

\begin{thm} \la{Same_fq}
Assume all oscillators $j \in \Sigma_N$ share the same frequency, $\omega_j = \Omega$, and phase-lock. Then $\dot{|r|} = 0, \, r(t) = |r| e^{i\Omega t}$ and 
$z_j(t) \in \{e^{i \Omega t}, \, -e^{i \Omega t} \}, \, \forall j \in \Sigma_N$, up to an overall fixed arbitrary phase. 
\end{thm}

{\it{Proof. }} Obviously, the trivial special case $r = 0$ is omitted. From Theorem~\ref{Phase-locking conditions}, we find that the positions of all the oscillators can only be at  two points, $\frac{zr}{|r|}, -\frac{\bar{z}r}{|r|} \in S^1$. However, by convexity, the phase of $r$ should lie between these two points, which is a contradiction unless $z = 1$, so $ z_j \in \left \{\frac{r}{|r|}, -\frac{r}{|r|}  \right \}$, and the value of $|r|$ is therefore given by

$$
|r| = \frac{1}{N} 
\left | 
\#  \left \{z_j = \frac{r}{|r|} \right  \} - \#  \left \{z_k = - \frac{r}{|r|} \right  \}
\right |, 
$$
 \noindent such that  $\dot{|r|}=0, \, [r, z_j] = 0$, and the Kuramoto equations become 
$$
\dot r = i r \Omega, 
$$

\noindent which completes the proof.  \hfill $\square$

This result allows us to introduce the last group of important definitions for the Kuramoto system, providing the sufficient condition counterpart for Theorem~\ref{Same_fq}:

\begin{definition}
A system of Kuramoto oscillators is called {\it{unsynchronized}} if $|r| = 0$, {\it{(partially) synchronized}} if $|r| \in (0, 1)$ is a constant, and {\it{fully synchronized}} if $|r| = 1$. 
\end{definition}

\begin{thm} 
Assume a system of Kuramoto oscillators with identical natural frequencies $\omega_j = \Omega, \, \forall j \in \Sigma_N$ and positive coupling constant, is synchronized. Then the system is phase-locked, as described by Theorem~\ref{Same_fq}. 
\end{thm}

{\it{Proof. }} 
From Eq.~\ref{eq_11} with $\omega_j = \Omega, \, \forall j \in \Sigma_N$, we obtain by adding all the equations 

$$
\dot{r} = i \Omega r + \frac{i \lambda}{N} \sum_{j = 1}^N z_j [r, z_j] 
$$

\noindent On the other hand, $|r| = $ constant implies

$$
\dot{r} = i \dot{\theta_0} r
$$

Therefore, since $|r| > 0$, subtracting these two equations yields 

$$
\dot \theta_0 - \Omega =  \frac{\lambda}{N} \sum_{j = 1}^N \frac{z_j}{r}  [r, z_j] \Rightarrow
\Omega - \dot \theta_0  =  \frac{\lambda}{N} \sum_{j = 1}^N e^{i \varphi_j} [e^{i (\theta_0 + \varphi_j)}, e^{i \theta_0}] = 
 \frac{\lambda}{N} \sum_{j = 1}^N \sin(\varphi_j) e^{i \varphi_j}
$$

Taking the imaginary part, we find, at $\lambda > 0$, 

$$
0 = \sum_{j=1}^N \sin^2(\varphi_j) \Rightarrow \varphi_j \in \{ 0, \pi \}, \, \forall j \in \Sigma_N
$$

Therefore, the oscillators are all phase-locked, and Theorem~\ref{Same_fq} applies. Furthermore, from the real part, we obtain

$$
\Omega - \dot \theta_0  =   \frac{\lambda}{2N} \sum_{j = 1}^N \sin(2 \varphi_j) = 0 \Rightarrow \dot \theta_0 = \Omega. 
$$ \hfill $\square$

\section{A generalized Kuramoto model as a mean-field spin system in $\mathbb{R}^3$} \la{second}

In the remaining sections of the article, we will use the notation $\vec{S}_j \in \mathbb{R}^3, \,\, j = 1, 2, \ldots, N$ to denote vectors in $\mathbb{R}^3$, generalizing the classical Kuramoto model. We will refer to these variables as (classical) {\it{spins}}. The mean field (their vector average, restricted to their convex hull), is denoted  

$$
\vec{J} := \frac{1}{N} \sum_{k=1}^N \vec{S}_k \in \mathbb{R}^3 %\mathbb{B}_3
$$

\begin{thm} \la{thm_extended}
The equations of motions for a system of spins $\{ \vec{S}_j \}, \, j \in \Sigma_N$, 

\be \la{eq_12}
\dot{\vec{S}}_j = \omega_j \widehat{e}_3 \times \vec{S}_j +  \lambda  \vec{S}_j \times  (\vec{J} \times \vec{S}_j )
\ee
\noindent reduce to the Kuramoto equations Eq.~\ref{eq_11} when the spins are normalized $|\vec{S}_j = 1|$, co-planar in the plane perpendicular to $\widehat{e}_3$, $\vec{S}_j(t) = %\langle z_j(t), 0 \rangle = 
\langle \cos \theta_j,(t) \sin \theta_j(t), 0 \rangle,  \, \forall \,  j \in \Sigma_N$.
\end{thm}

\noindent {\it{Proof. }} Let us first remark that Eqs.~\ref{eq_12} imply $\vec{S}_j \cdot \dot{\vec{S}}_j = 0,$ so the norms $|\vec{S}_j|$ are preserved by this dynamics.  If the vectors also have the unitary, co-planar initial conditions $\vec{S}_j(t=0) = \langle z_j, 0 \rangle = \langle \cos \theta_j, \sin \theta_j, 0 \rangle, \,\, j \in \Sigma_N$, then  $\vec{J} \times \vec{S}_j \parallel \widehat{e}_3$, so $\dot {\vec{S}_j} \perp \widehat{e}_3$, $\widehat{e}_3 \cdot \dot{\vec{S}}_j = 0$, and  the vectors remain co-planar, unit vectors for $t > 0$.  In this case, identifying 

$$
\dot{\vec{S}}_j = \dot{\theta}_j \widehat{e}_3 \times \vec{S}_j, \quad 
\vec{J} \times \vec{S}_j = -\left [ \frac{1}{N}\sum_{k=1}^N \sin(\theta_k - \theta_j) \right ] \widehat{e}_3, 
$$

\noindent Eqs.~\ref{eq_12} become

$$
 \dot{\theta}_j \widehat{e}_3 \times \vec{S}_j = \left \{ 
 \omega_j + \frac{\lambda}{N}\sum_{k=1}^N \sin(\theta_k - \theta_j)
 \right \}
 \widehat{e}_3 \times \vec{S}_j,
$$
equivalent to Eqs.~\ref{eq_1} and Eqs.~\ref{eq_11}.  \hfill $\square$ 

\subsection{Global properties of the generalized Kuramoto model in $\mathbb{R}^3$} \la{Properties}

\subsubsection{The global synchronization condition and asymptotic states.}

\noindent Defining $\vec{P}_j : = \vec{S}_j \times \widehat{e}_3$, such that $\vec{S}_j, \, \vec{P}_j$ are dual to each other when perpendicular to $\widehat{e}_3$, imposing that all the spins be co-planar, multiplying in Eqs.~\ref{eq_12} by $\vec{J}$ and summing over $j$ leads to

$$
\frac{1}{2}\frac{d \,\,}{dt} |\vec{J}|^2 = - \sum_{j=1}^N \omega_j (\vec{P}_j \cdot \vec{J}) + \lambda \sum_{j=1}^N (\vec{P}_j \cdot \vec{J})^2
$$

\noindent Therefore, the system reaches a constant value for the magnitude of the order parameter $|\vec{J}|$ when

$$
\sum_{j=1}^N \omega_j (\vec{P}_j \cdot \vec{J}) = \lambda \sum_{j=1}^N (\vec{P}_j \cdot \vec{J})^2,
$$

\noindent or by using the phase difference $\varphi_j$ between $\vec{S}_j$ and $\vec{J}$, as well as $\sum_{j=1}^N \sin(\varphi_j) = 0$,

%$$
\be \la{global}
\lambda |\vec{J}| = \frac{\sum_{j=1}^N \omega_j \sin(\varphi_j)}{\sum_{j=1}^N \sin^2(\varphi_j)} 
=
 \frac{\sum_{j=1}^N (\omega_j - \Omega) \sin(\varphi_j)}{\sum_{j=1}^N \sin^2(\varphi_j)} 
\ee
%$$

Clearly, for a generic distribution of natural frequencies $\{ \omega_j \}_{j=1}^N$, complete synchronization corresponds to $|\vec{J}| \to 1^{-}$, $\varphi_j \to 0, \, \forall j \in \Sigma_N$, such that $\lambda |\vec{J}| \to \infty$, so it can only occur in the $\lambda \to \infty$ limit.

Writing the equations of motion in the form

$$
\dot{\theta}_j = \omega_j - \lambda \widehat{e}_3 \cdot (\vec{J} \times \vec{S}_j), 
$$

\noindent the phase-locking conditions for spins $j, k$ become

$$
\omega_j - \omega_k = \lambda |\vec{J}| (\sin \varphi_j - \sin \varphi_k) \Rightarrow 
\frac{\sin \varphi_j - \sin \varphi_k}{\omega_j - \omega_k} = \frac{1}{\lambda |\vec{J}|}
$$

\noindent In the case where the coupling constant is large enough so that $|\omega_j - \Omega |\le \lambda |\vec{J}|, \, \forall j \in \Sigma_N$, $\Omega = \frac{1}{N} \sum_k \omega_k$, a solution for the system of linear equations for variables $\sin \varphi_k $ is

\be \la{solution}
\sin \varphi_j = \frac{\omega_j - \Omega}{\lambda |\vec{J}|}, \quad {\rm{sign}} (\cos \varphi_j) := \epsilon_j \in \{-1, 1\}, 
\ee

\noindent consistent with the general synchronization condition Eq.~\ref{global} found earlier and the 3-spin particular solution, Eq.~\ref{sol_3}, with $\alpha^{-1} = \lambda |\vec{J}|, \, \beta = \Omega$. Since

\be \la{critical}
%$$
|\vec{J}| = \frac{1}{N} \sum_{j=1}^N \cos \varphi_j = \frac{1}{N}  \sum_{j=1}^N \epsilon_j \sqrt{1 - \frac{(\omega_j - \Omega)^2}{\lambda^2 |\vec{J}|^2}}, \quad 
{\rm{where}} \quad \epsilon_j = \pm 1,
%$$
\ee

\noindent we arrive at an exact formula for the dependence $|\vec{J}| (\lambda)$, in the complete synchronization regime $\lambda \ge \lambda_s$, but without the asymptotic assumption $\lambda \to \infty, \, |\vec{J}| \to 1$. 

Without loss of generality, let $N_+  \ge N_-$ represent the numbers of oscillators for which $\epsilon_j = 1$ and $-1$, respectively, such that $N_+ + N_- = N$. Assuming that $\lambda$ is large enough such that $N |\omega_j - \Omega |\ll \lambda (N_+ - N_-), \, \forall j \in \Sigma_N$, we can expand the solution to obtain the asymptotic equilibria approximation 

$$
|\vec{J}| \approx \frac{N_+ - N_-}{N} - \frac{N^2}{2\lambda^2 (N_+ - N_-)^2} \left [\frac{1}{N}\sum_{j=1}^N \epsilon_j  (\omega_j - \Omega)^2\right ] \approx  |\vec{J}| _0 - \frac{\mathbb{E}[\epsilon_j (\omega_j - \Omega)^2]}{2 \lambda^2 |\vec{J}|^2 _0}, 
$$

\noindent where 

$$
|\vec{J}| _0 = \frac{N_+ - N_ -}{N} 
$$

\noindent is the $\lambda \to \infty$ limit of the solution with fixed $N_+, N_-$. Therefore, the large coupling constant limit allows for a whole family of solutions in which individual spins are clustered around two opposite phases, and angular spread proportional to the frequency deviation from the average frequency.  

Consistent with Theorem~\ref{Same_fq}, in the case of equal frequencies, $\omega_j = \Omega$,  the equilibrium condition becomes

$$
\lambda |\vec{J}| \sum_{j=1}^N \sin^2(\varphi_j) = 0,
$$

\noindent which can be satisfied as $\lambda = 0$ (trivial), $\vec{J} = \vec{0}$ (unsynchronized), or $\sum_{j=1}^N \sin^2(\varphi_j) = 0$ (synchronized), corresponding to some fixed value $|\vec{J}|_0 = \frac{N_+ - N_-}{N}$. 

\noindent The collective spin dynamics towards this globally-stable synchronization is given by

$$
%\frac{1}{2|\vec{J}|}\frac{d \,\,}{dt} |\vec{J}|^2 =
\frac{d \,\,}{dt} |\vec{J}|
= \lambda \sum_{j=1}^N \sin^2(\varphi_j) \ge 0, 
$$

\noindent so the order parameter will monotonically increase towards its equilibrium value, reached when $\varphi_j \in \{ 0, \pi \}, \, \forall j \in \Sigma_N$. 

\subsubsection{Spin-flip dynamics on a stationary background.} \la{SF}  The generic solutions found in the previous section are not globally stable, as perturbations of one oscillator with $\epsilon_j = -1$ will lead to the spin-flip change $N_+ \to N_+ + 1, \, N_- \to N_- -1$.  This dynamics has a kink-like solution in the limit $N \to \infty$, since setting the background total spin to the stationary evolution 

$$
\frac{d \vec{J}}{dt}  = \Omega  \, \widehat{e}_3 \times \vec{J}, \quad |\vec{J}| = \frac{N_+ - N_-}{N}, 
$$

\noindent and denoting by $\omega, \, \varphi$ the natural frequency and angle difference between the oscillator with $\epsilon_j = -1$ and the vector $\vec{J}$, the solitary spin dynamics solves

$$
\dot \varphi = \omega - \Omega - \lambda |\vec{J}| \sin \varphi, 
$$

\noindent describing the evolution from the unstable solution $\varphi_*$ towards the stable solution $\varphi_0$, 

\be \la{1-spin}
\sin \varphi_* = \frac{\omega - \Omega}{\lambda |\vec{J}|}, \,\, \cos \varphi_* < 0 \longrightarrow 
\sin \varphi_0 = \frac{\omega - \Omega}{\lambda |\vec{J}|}, \,\, \cos \varphi_0 > 0.
\ee

\noindent Stability of this solution can be verified in the small perturbation approximation $\varphi = \varphi_0 + \delta, \, |\delta| \ll 1$, 

$$
\dot \delta =  \omega - \Omega - \lambda |\vec{J}| \sin(\varphi_0 + \delta) = 
(\omega - \Omega)(1-\cos \delta) - \lambda |\vec{J}| \cos(\varphi_0) \, \sin \delta ,
$$ 

\noindent therefore, since $ \lambda |\vec{J}|  \cos \varphi_0 =  \sqrt{(\lambda |\vec{J}|)^2 - (\omega - \Omega)^2} > 0$,  reducing in the limit $\delta \to 0$ to 

$$
\dot \delta \approx \sqrt{(\lambda |\vec{J}|)^2 - (\omega - \Omega)^2} \, \sin \delta,  
$$

\noindent whose solution, 

\be \la{relax}
\delta(t) = 2 \arctan \left [\tan \left (\frac{\delta(0)}{2} \right ) e^{- \Lambda_{\omega} t} \right ], \quad \Lambda_{\omega} = \sqrt{(\lambda |\vec{J}|)^2 - (\omega - \Omega)^2} \,\,, 
\ee

\noindent is a 0-D sine-Gordon 1-soliton solution and describes a kink spin-flip (in the case $\omega = \Omega$, when the phase change equals exactly $\pi$, and the topological charge changes by one unit, $N_+ \to N_+ + 1, \, N_- \to N_- -1$).  

For the case of two spins $\varphi_{1, 2}$ undergoing the phase flip, their collective dynamics is given by the system of equations 

\be \la{2-spins}
\left \{
\begin{array}{lcl}
\dot \varphi_1 &  = &  \omega_1 - \Omega - \lambda J \sin \varphi_1 - \frac{\lambda}{N}\sin (\varphi_1 - \varphi_2) \\
 & & \\
\dot \varphi_2 &  = &  \omega_2 - \Omega - \lambda J \sin \varphi_2 - \frac{\lambda}{N}\sin (\varphi_2 - \varphi_1) 
\end{array}
\right .
\ee

\noindent Introducing the symmetric $\sigma = \frac{1}{2}(\varphi_1 + \varphi_2)$ and antisymmetric $\delta = \frac{1}{2}(\varphi_1 - \varphi_2)$ combinations, the system of equations Eqs.~\ref{2-spins} becomes 

$$
\left \{
\begin{array}{lcl}
\dot \sigma &  = &  \left ( \frac{\omega_1 + \omega_2}{2}  - \Omega\right ) - \lambda J \cos(\delta) \sin (\sigma)  \\
 & & \\
\dot \delta &  = &  \frac{\omega_1-\omega_2}{2} - \lambda J \sin(\delta) \cos (\sigma) - \frac{\lambda}{N} \sin \delta,
\end{array}
\right .
$$

\noindent where $J$ denotes the constant magnitude of the spin average of the other $N-2$ spins, rescaled by a factor of $\frac{N-2}{N}$. In the large $N$ limit, these equations simplify and their linearization has the asymptotic solution

$$
\sigma \approx \frac{\omega_1 + \omega_ 2 -  2\Omega}{2\lambda J}, \quad 
\delta \approx \frac{\omega_1-\omega_2}{2\sqrt{(\lambda J)^2 - \left [ 
\frac{\omega_1+\omega_2}{2} - \Omega 
\right ]^2}}, 
$$
\noindent consistent with 1-spin solutions Eq.~\ref{1-spin} up to corrections of the order $\frac{1}{(\lambda J )^2} \left [ \frac{\omega_1+\omega_2}{2} - \Omega \right ]^2$.  

\subsubsection{The complete synchronization limit.}

In the  limit $\sqrt{{\rm{Var}}(\omega)} \ll \lambda \to \infty, \, |\vec{J}| \to 1$, we obtain the global synchronization approximation corresponding to $\epsilon_j = 1, \, \forall j \in \Sigma_N$, 

$$
|\vec{J}| \approx 1 - \frac{1}{2\lambda^2 N |\vec{J}|^2} \sum_{j=1}^N (\omega_j - \Omega)^2 \approx  1 - \frac{{\rm{Var}}(\omega)}{2 \lambda^2}
\approx \sqrt{\frac{\lambda^2 - {{\rm{Var}}(\omega)}}{\lambda^2}}
$$

\noindent This solution is globally stable as no further spin-flip dynamics is possible. Individual spins relax back to their equilibrium phases exponentially fast, with the decay rate  $\Lambda_{\omega}$ found in Eq.~\ref{relax}.  Asymptotically $(\sqrt{{\rm{Var}}(\omega)} \ll \lambda \to \infty)$, the limiting rates are $\Lambda_{\omega} \to \lambda$.

\subsection{Lagrangian structure of the generalized Kuramoto model.}

So far, all the known characteristics of the Kuramoto system have been confirmed in this generalized model. Therefore, it is  now justified to consider the Lagrangian and Hamiltonian structures compatible with the generalized Kuramoto equations Eqs.~\ref{eq_12}. 

\begin{thm} \la{Thm_Lagrangian} 
Define the Lagrangian with configuration space $\mathbb{R}^{3N}$ 
%\be \la{eq_13}
$$
L(\{\vec{S}_j\}, \{\dot{\vec{S}}_j \}) = 
\sum_{j=1}^N \left \{
\widehat{e}_3 \cdot( {\dot{\vec{S}}}_j \times \vec{S}_j ) 
- \omega_j |\vec{S}_j|^2 +
\lambda \left ( 
\widehat{e}_3 \times \left [
( \vec{J}\times \vec{S}_j)  \times  \vec{S}_j
\right ]
\right ) \cdot \vec{S}_j
\right \}
$$
%\ee
Equations \ref{eq_12} are the Euler-Lagrange equations for the action 
$$
W[\{\vec{S}_j, \dot{\vec{S}}_j \}_{j\in \Sigma_N}] := \int_0^T L(\{\vec{S}_j\}, \{\dot{\vec{S}}_j \})  dt
$$
\end{thm} 

{\it{Proof. }} Denote by 
$$
\Lambda_1 := \widehat{e}_3 \cdot( {\dot{\vec{S}}}_j \times \vec{S}_j ), \quad
\Lambda_2 :=  -\omega_j |\vec{S}_j|^2, \quad 
 \Lambda_3 := \lambda \left [ 
\widehat{e}_3 \times \left (
( \vec{J}\times \vec{S}_j)  \times  \vec{S}_j
\right )
\right ]\cdot \vec{S}_j, 
$$
\noindent such that $L$ consists of a sum over all spins of the terms $\Lambda_1 + \Lambda_2 + \Lambda_3,$ and compute 
$$
\frac{\p \Lambda_1}{\p (\dot{\vec{S}}_j)_{\alpha}} = %\sum_{j=1}^N 
\frac{\p \,\,\, }{\p (\dot{\vec{S}}_j)_{\alpha}}  \epsilon_{3\gamma\beta}(\vec{S_j})_\beta (\dot{\vec{S}}_j)_\gamma 
=  \epsilon_{3\alpha\beta}(\vec{S_j})_\beta = - (\widehat{e}_3 \times \vec{S}_j)_\alpha
\Rightarrow 
$$
$$
\frac{\p \Lambda_1}{\p \dot{\vec{S}}_j} =-  \widehat{e}_3 \times \vec{S}_j, \quad 
\frac{\p \Lambda_1}{\p {\vec{S}}_j} =  \widehat{e}_3 \times \dot{\vec{S}}_j, \quad 
\frac{d \,}{dt} \left (\frac{\p \Lambda_1}{\p \dot{\vec{S}}_j}\right ) - \frac{\p \Lambda_1}{\p {\vec{S}}_j} = 
- 2 \,\widehat{e}_3 \times \dot{\vec{S}}_j,
$$
$$
\frac{\p \Lambda_2}{\p {\vec{S}}_j} = -2 \, \omega_j \vec{S}_j, 
$$
$$
\Lambda_3 = \lambda \left \{
\widehat{e}_3 \times \left [
(\vec{J}\cdot \vec{S}_j) \vec{S}_j - |\vec{S}_j|^2 \vec{J}
\right ]
\right \} 
\cdot \vec{S}_j = 
- \lambda |\vec{S}_j|^2  \left [
\widehat{e}_3 \times \left (
\vec{J} - \frac{1}{N}\vec{S}_j 
\right )
\right ]\cdot \vec{S}_j,
$$
\noindent using properties of vector products. Since $\vec{J} - \frac{1}{N}\vec{S}_j $ does not depend on $\vec{S}_j$, we can compute 
$$
\frac{\p \Lambda_3}{\p {\vec{S}}_j} = 
-2 \lambda \left ( \left [
\widehat{e}_3 \times \left (
\vec{J} - \frac{1}{N}\vec{S}_j 
\right )
\right ]\cdot \vec{S}_j \right ) \vec{S}_j - 
\lambda |\vec{S}_j|^2 
\widehat{e}_3 \times \left (
\vec{J} - \frac{1}{N}\vec{S}_j 
\right )
$$
$$
\frac{\p \Lambda_3}{\p {\vec{S}}_j} = 
-2 \lambda  \left [
\left (
\vec{J} 
 \times 
 \vec{S}_j
\right )
\cdot\widehat{e}_3 \right ] \vec{S}_j - 
\lambda |\vec{S}_j|^2 
\widehat{e}_3 \times \left (
\vec{J} - \frac{1}{N}\vec{S}_j 
\right ) : =
\Lambda_3^{(0)} + \Lambda_3^{(1)}. 
$$

Taking separately the combination

$$
\frac{d \,}{dt} \left (\frac{\p \Lambda_1}{\p \dot{\vec{S}}_j}\right ) - \frac{\p \Lambda_1}{\p {\vec{S}}_j} 
- \frac{\p \Lambda_2}{\p {\vec{S}}_j}  
- \Lambda_3^{(0)} = 
- 2 \,\widehat{e}_3 \times \dot{\vec{S}}_j + 2 \, \omega_j \vec{S}_j + 2 \lambda  \left [
\left (
\vec{J} 
 \times 
 \vec{S}_j
\right )
\cdot\widehat{e}_3 \right ] \vec{S}_j,  
$$

\noindent and multiplying from the left by $\widehat{e}_3 \times $ yields 

$$
2 \left \{
(\widehat{e}_3 \cdot \dot{\vec{S}}_j) \widehat{e}_3 - \dot{\vec{S}}_j 
+ \omega_j \, \widehat{e}_3 \times \vec{S}_j + 
\lambda  \left [
\left (
\vec{J} 
 \times 
 \vec{S}_j
\right )
\cdot\widehat{e}_3 \right ] \,  \widehat{e}_3 \times \vec{S}_j
\right \}
$$

Denoting by 
$$
\widehat{P}_{3} (.) := (\widehat{e}_3 \cdot \, \, . \,\,) \widehat{e}_3
$$

\noindent the projector onto the $\widehat{e}_3$ subspace, the expression becomes

$$
2 \left \{
 \widehat{P}_3 \dot{\vec{S}}_j  - \dot{\vec{S}}_j 
+ \omega_j \, \widehat{e}_3 \times \vec{S}_j + 
\lambda   \vec{S}_j \times \left [ \widehat{P}_3
\left (
\vec{J} 
 \times 
 \vec{S}_j
\right )
\right ] \, 
\right \}
$$

Setting the expression to zero, we obtain the equations

$$
 \dot{\vec{S}}_j -  \widehat{P}_3 \dot{\vec{S}}_j  = 
 \omega_j \, \widehat{e}_3 \times \vec{S}_j + 
\lambda  \vec{S}_j\times  \left [ \widehat{P}_3
\left (
\vec{J} 
 \times 
 \vec{S}_j
\right )
\right ] \, 
$$

Clearly, for the case of planar spins, $ \widehat{P}_3 \dot{\vec{S}}_j = \vec{0}$ and $ \widehat{P}_3
\left (
\vec{J} 
 \times 
 \vec{S}_j
\right ) = \vec{J} 
 \times 
 \vec{S}_j$, yielding equations Eqs.~\ref{eq_12} and therefore the Kuramoto equations for the planar spins case. 

Finally, to compensate for the sum of the terms $\Lambda_3^{(1)}$, which equals
$$
-\lambda \sum_{j=1}^N \widehat{e}_3 \times \left ( \vec{J} - \frac{1}{N}\vec{S}_j \right ) = -\lambda \frac{N-1}{N}\, \widehat{e}_3 \times \vec{J}
=  -\lambda \frac{N-1}{N^2} \sum_{j=1}^N  \, \widehat{e}_3 \times \vec{S}_j
$$

\noindent we notice that 

$$
\frac{\p \vec{S}_j \cdot (\widehat{e}_3 \times \vec{S}_j)}{\p \vec{S}_j} = 
2\widehat{e}_3 \times \vec{S}_j, 
$$

\noindent so adding to the Lagrangian the term 

$$
\lambda \frac{N-1}{2N^2} \sum_{j=1}^N \vec{S}_j \cdot (\widehat{e}_3 \times \vec{S}_j)
$$

\noindent would indeed cancel out the sum of terms $\Lambda_3^{(1)}$. However, since 

$$
\vec{S}_j \cdot (\widehat{e}_3 \times \vec{S}_j) = \widehat{e}_3  \cdot (\vec{S}_j \times \vec{S}_j)  \equiv 0,
$$

\noindent this means that the action $W$ is sufficient to generate the equations Eqs.~\ref{eq_12}.  \hfill $\square$ 

\subsubsection{A Hamiltonian structure for the generalized Kuramoto model}

Since 

$$
\frac{\p L}{\p \dot{\vec{S}}_j} =-  \widehat{e}_3 \times \vec{S}_j = \vec{S}_j \times  \widehat{e}_3 := \vec{P}_j 
$$

\noindent defines the associated momentum variable $\vec{P}_j$, and the system's Hamiltonian function is the Legendre transform 
$H = \sum_{j=1}^N \vec{P}_j \cdot \dot{\vec{S}}_j - L$,  we seek to define a function on the phase space, symmetric with respect to the pairs of variables $\{\vec{S}_j, \vec{P}_j \}, \, j \in \Sigma_N$. The following choice can be seen as a natural candidate:

\begin{thm}\la{Thm_Hamiltonian}
The generalized Kuramoto model admits the Hamiltonian
$$
H(\{\vec{S}_j, \vec{P}_j\}) = 
\sum_{j=1}^N \left \{
 -\frac{\omega_j}{2}   \left [|\vec{S}_j|^2 + 
 |\vec{P}_j|^2 
\right ]
+
\lambda (\vec{J}\times \vec{S}_j)\cdot ( \vec{P}_j \times \vec{S}_j)
%+
%\lambda (\vec{K}\times \vec{P}_j)\cdot ( \vec{S}_j \times \vec{P}_j)
%
\right  \}, 
%  -
%\lambda \left [ 
%\widehat{e}_3 \times \left (
%( \vec{J}\times \vec{S}_j)  \times  \vec{S}_j
%\right )
%\right ]\cdot \vec{S}_j
%\right \}
$$
\noindent under planar Kuramoto initial conditions. 
\end{thm}

%\noindent 

{\it{Proof.}} With respect to its independent variables, we have the Hamiltonian's derivatives 

$$
\frac{\p  H}{\p \vec{P}_j} = -\omega_j \vec{P}_j + \lambda \vec{S}_j \times (\vec{J} \times \vec{S}_j) =  \omega_j \widehat{e}_3 \times \vec{S}_j + \lambda \vec{S}_j \times (\vec{J} \times \vec{S}_j), 
$$

\noindent which indeed equal $\dot{\vec{S}_j}$ as in Eqs.~\ref{eq_12}. For the second set of equations, we have

$$
\frac{\p  H}{\p \vec{S}_j} =  -\omega_j \vec{S}_j  +  \lambda  (\vec{J} \times \vec{S}_j) \times \vec{P}_j +
\lambda  (\vec{P}_j \times \vec{S}_j) \times \vec{J} - 
\frac{\lambda}{N} \sum_{k \ne j} \vec{S}_k \times (\vec{P}_k \times \vec{S}_k)
$$

\noindent The last term equals $\lambda \widehat{e}_3 \times \vec{J}$ in the planar spins configuration, so we need to compare
$$
- \dot{\vec{P}}_j + \omega_j \vec{S}_j= \widehat{e}_3 \times \dot{\vec{S}}_j +  \omega_j \vec{S}_j = \lambda \widehat{e}_3 \times[ \vec{S}_j \times (\vec{J} \times \vec{S}_j)] 
=
\lambda \widehat{e}_3 \times \vec{J} 
+
\lambda (\vec{J}\cdot \vec{S}_j) \vec{P}_j
$$

\noindent to 

$$
\lambda  (\vec{J} \times \vec{S}_j) \times \vec{P}_j +
\lambda  (\vec{P}_j \times \vec{S}_j) \times \vec{J} +
\lambda \widehat{e}_3 \times \vec{J},
$$

\noindent or

$$
  (\vec{P}_j \times \vec{S}_j) \times \vec{J} =  \widehat{e}_3 \times \vec{J}
$$

\noindent and 

$$
- (\vec{J}\cdot \vec{S}_j) \vec{P}_j + (\vec{J} \times \vec{S}_j) \times \vec{P}_j  =
- (\vec{J}\cdot \vec{S}_j) \vec{P}_j  + (\vec{J}\cdot \vec{P}_j) \vec{S}_j 
$$

\noindent For any pair of canonically-oriented, mutually orthogonal planar unit vectors $ \vec{P}_j, \vec{S}_j$ this identiy holds, so indeed with planar initial conditions, the Hamilton equations for the Hamiltonian in Theorem~\ref{Thm_Hamiltonian} become the (self-dual) Kuramoto equations for the spin variables $\{ \vec{S}_j \}, \, j \in \Sigma_N$.  \hfill $\square$

%Furthermore, expanding the second term and summing over $j$ yields the simplified, final mean-field  form
%
%\be \la{eq_Hamiltonian}
%H(\{\vec{S}_j, \vec{P}_j\}) = 
%\sum_{j=1}^N \left \{
% \omega_j |\vec{S}_j|^2 
% +
%\lambda (\vec{J}\times \vec{S}_j)\cdot ( \vec{P}_j \times \vec{S}_j)
%\right \}
%\ee

%\noindent 
Notice that, as expected, the planar reduction of the Hamiltonian yields the constant value (total energy) 

$$
E =  - \sum_{j=1}^N \omega_j. 
$$

%Using the orthogonal decomposition
%
%$$
%\vec{S}_j = \widehat{P}_3 (\vec{S}_j ) - \widehat{e}_3 \times \vec{P}_j ,  
%$$
%
%\noindent we have the equivalent form
%
%$$
%H = \sum_{j=1}^N \left \{
% \omega_j \left (
% |\vec{P}_j|^2 + |\widehat{P}_3 (\vec{S}_j )|^2
% \right )
% +
%\lambda (\vec{J}\times \vec{S}_j)\cdot ( \vec{P}_j \times \vec{S}_j)
%\right \}
%$$
%
%In these variables, Eqs.~\ref{eq_12} take the form
%
%$$
%\dot{\vec{S}}_j = - \omega_j \vec{P}_j + \lambda [\vec{J} -(\vec{S}_j \cdot \vec{J})\vec{S}_j] =
% - \omega_j \vec{P}_j + \lambda \left [\mathbb{I} - \widehat{P}_{\vec{S}_j} \right ] \vec{J}. 
%$$

\subsubsection{Planar perturbation Hamiltonian around general synchronization solutions.} As seen in \S~\ref{Properties}, for sufficiently large values of the coupling constant, the system may reach synchronization configurations $\{\vec{S}_j^{(0)} \}_{j \in \Sigma_N}$ Eqs.~\ref{solution}, which can already be classified by the numbers of spins in each sector $(N_+, N_-)$. Considering a small in-plane perturbation of the Hamiltonian in Theorem~\ref{Thm_Hamiltonian} around such a solution, we introduce the notation 

$
\vec{S}_j \approx \vec{S}_j^{(0)} + \vec{\sigma}_j, \,\, \vec{\sigma}_j\cdot \vec{S}_j^{(0)} = 0, \quad 
\vec{\sigma}_j = |\vec{\sigma}_j| \vec{S}_j^{(0)}  \times \widehat{e}_3,  %\quad \sum_{j=1}^N
$

\noindent such that 

$$
H \approx H_0(\{\vec{S}_j^{(0)} \}_{j \in \Sigma_N}) + h(\{\vec{\sigma}_j \}_{j \in \Sigma_N}), \quad
h = - \lambda \sum_{j=1}^N \vec{J}\cdot \vec{\sigma}_j^*, 
$$

\noindent with pseudo-spins $\vec{\sigma}_j^*$ given by the vectors dual to $\vec{\sigma}_j$. The first-order planar perturbation Hamiltonian is therefore typical for a Heisenberg spin model, its minimum value at fixed $|\vec{J}|$ given by the complete synchronization solution corresponding to a ferromagnetic ground state in a spin model with Hamitonian $h$,
$$
\vec{J}\cdot \vec{\sigma}_j^* > 0, \,\, \forall j \in \Sigma_N.
$$

%\section{Comparing the Kuramoto model to the semiclassical Gaudin magnet} \la{third}
%
%Consider now a Kuramoto synchronized state Eqs.~\ref{solution}, such that 
%
%$$
%\vec{S}_j \in S^1, \quad \lambda \vec{J} \times \vec{S}_j = (\omega_j - \Omega)\widehat{e}_3, \quad \vec{P}_j \times \vec{S}_j = \widehat{e}_3, \quad \forall \, j \in \Sigma_N. 
%$$
%
%\noindent We emphasize that the equilibrium configuration does not need to be stable that is $N_+ \ne N_-$. Consider an off-plane perturbation of this state, $\vec{S}_j \to \vec{S}_j + \sigma_j \widehat{e}_3$, and identify $\sigma_j = S_j^{(3)}$, the axial component of the spin variable in $\mathbb{R}^3$. Expanding,
%
%$$
% \lambda \vec{J} \times \vec{S}_j \approx  (\omega_j - \Omega)\widehat{e}_3 + \lambda S_j^{(3)} \vec{J}^{(\perp)} \times \widehat{e}_3 -   \lambda  \vec{J}^{(\perp)} \times \vec{S}_j^{(\perp)}, 
%$$
%
%\noindent where $\vec{J}^{(\perp)}$ is the in-plane component of the mean-field spin, and 
%
%$$
%\vec{P}_j \times \vec{S}_j \to \widehat{e}_3 
%$$
 
 \section{Geometric quantization of the generalized Kuramoto system around synchronization equilibria}
 
 \subsection{Consistency of classical Poison brackets and their deformation quantization} 
 
 Notice that the formal Poisson bracket structure found in the previous section for the generalized Kuramoto model in Theorem~\ref{Thm_Hamiltonian} is: 
 
 $$
 \frac{d f}{dt} = \{ H, f \} := \sum_{j=1}^N 
 \frac{\p H}{\p \vec{P}_j} \cdot \frac{\p f}{\p \vec{S}_j} - 
 \frac{\p f}{\p \vec{P}_j} \cdot \frac{\p H}{\p \vec{S}_j},
 $$
 
 \noindent for any function depending on time only via the variables $\{\vec{S}_j, \, \vec{P}_j \}_{j \in \Sigma_N}$. This formally extends to what may at first appear to be vector variables, such as $\vec{P}_k, \, \vec{S}_{\ell}$, whereas 
 
 $$
  \{ \vec{P}_j, \vec{S}_k \} = \delta_{jk}, \,\, j, k \in \Sigma_N. 
 $$
 
 That this is, indeed, a proper Poisson bracket structure for functions of the planar limit of the generalized Kuramoto model follows from the fact that, while constrained to the unit circle, variables $\vec{P}_k, \, \vec{S}_{\ell}$ are, in fact, scalars (equivalent to their angular variables, more precisely). In fact, it is possible to identify this choice of Poisson brackets to the antisymmetric bilinear form defined in Eqs.~\ref{eq_10}, as indeed for unimodular complex variables $u = e^{i \alpha}, \, v = e^{i \beta}, \, \alpha, \beta \in \mathbb{R}$, we have
 
 $$
 [u, v] = \sin(\alpha - \beta) = \widehat{e}_3 \cdot (\vec{v} \times \vec{u}), 
 $$
 
 \noindent where $\vec{u}, \vec{v}$ are the planar vector representations $\vec{u} = \langle\cos \alpha, \sin \alpha, 0 \rangle, \, \vec{v} = \langle\cos \beta, \sin \beta, 0 \rangle$. Therefore, since $\vec{S}_j =  \langle\cos \theta_j, \sin \theta_j, 0 \rangle, \, \vec{P}_j =  \langle\cos (\theta_j - \frac{\pi}{2}), \sin (\theta_j - \frac{\pi}{2}), 0 \rangle,$ we can verify the identities 
 
 $$
   \{ \vec{P}_j, \vec{S}_j \} = 1 = \sin \left (\frac{\pi}{2} \right )  = \widehat{e}_3 \cdot (\vec{P}_j \times \vec{S}_j).
 $$
 
 \subsection{Quantization of off-planar fluctuations around Kuramoto synchronized states}
 
 Let us consider now an embedding of the Lie group $SU(1) \simeq S^1$, on which the classical variables $\{ \vec{S}_j, \, \vec{P}_j\}_{j\in \Sigma_N}$ are defined, and their minimal deformation into $SU(2) \simeq S^3$, as a double cover over the closed unit ball in $\mathbb{R}^3, \, \overline{\mathbb{B}}_3$.  Notice that, consistently, this will also contain all convex combinations of classical spins, such as the mean field $\vec{J}$. 
 
 Using the standard representation of $SU(2)$ and its algebra $\textswab{s:u}(2)$ through the Pauli matrices $\{\sigma_{\alpha} \}_{\alpha = 1, 2, 3}$, normalized such that 
$$
\sigma_{\alpha} \sigma_{\beta} = \frac{1}{4}\delta_{\alpha \beta} \mathbb{I} + \frac{i}{2} \epsilon_{\alpha \beta \gamma} \sigma_{\gamma}, 
\quad [\sigma_{\alpha}, \sigma_{\beta} ] = i  \epsilon_{\alpha \beta \gamma} \sigma_{\gamma}, 
$$  
\noindent where the canonical Lie bracket of $\textswab{s:u}(2)$ is used, we quantize the Kuramoto Hamiltonian system by deforming the classical Poisson brackets   of variables depending on time only through the canonical variables $\vec{P}_k, \, \vec{S}_{k}$ as 
$$
\widehat{e}_3 \cdot (\vec{a} \times \vec{b}) \to \{\vec{a}, \, \vec{b} \} + \hbar [\vec{a}\cdot \vec{\sigma}, \vec{b}\cdot \vec{\sigma}], 
$$ 
\noindent where $\vec{\sigma} := \langle \sigma_1, \sigma_2, \sigma_3 \rangle$. Expanding to first order in the deformation parameter $\hbar$ and using the identity
$$
 [\vec{a}\cdot \vec{\sigma}, \vec{b}\cdot \vec{\sigma}] = 2 (\vec{a}\cdot \vec{\sigma}) \cdot (\vec{b}\cdot \vec{\sigma}) - \frac{1}{2}(\vec{a}\cdot \vec{b})\, \mathbb{I}, 
$$
the semiclassical limit of this quantization will be obtained by replacing the operators with their ground-state averages, 
$$
\vec{a}\cdot \vec{\sigma} \to \vec{a}\cdot \langle\!\langle\vec{\sigma} \rangle\!\rangle := \vec{a}\cdot \vec{\tau}.
$$
\noindent For an off-plane perturbation, the ground state is an eigenvector of $\sigma_3$, so we have $\langle\!\langle \sigma_{\alpha} \rangle\!\rangle = 2 t_3 \delta_{\alpha 3}, \, \alpha = 1, 2, 3$, so the semiclassical limit reads 
$$
\vec{a}\cdot \vec{\sigma} \to 2a_3 t_3
$$

Collecting these results, the first-order perturbation (in $\hbar$) of the Kuramoto Hamiltonian $H \to H_0 + \hbar \widehat{H}_1$  around a synchronization state Eqs.~\ref{solution} becomes

$$
\widehat{H}_1 = \sum_{j=1}^N 2 (\omega_j - \Omega) t_j^{(3)} - \frac{\lambda}{N}|\vec{J}^{-}|^2, 
$$
\noindent where $\vec{J}^{-}: = \sum_{j=1}^N \langle t_j^{(1)}, t_j^{(2)}, 0 \rangle$ is the semiclassical limit of the planar total spin perturbation.  We notice that this coincides with the semiclassical limit of the Richardson-Gaudin Hamiltonian \cite{Dukelsky, Richardson1} describing the spin-pairing mechanism in an Anderson model with spectrum $\epsilon_j = \omega_j - \Omega, \, j \in \Sigma_N,$ and coupling constant $g = \frac{\lambda}{N}$.   

\subsection{Dynamics of the mean-field spin around Kuramoto synchronization states}

One of the main results that can be inferred from the analysis in the previous section is that the spin-flipping mechanism described perturbatively in \S~\ref{SF}, when the coupling constant is increased and the total classical spin is small ($\frac{N_+ - N_-}{N} = O \left ( \frac{1}{N}\right )$), is described by modulated elliptic  functions \cite{Dukelsky} and a separation of time scales roughly corresponding to $\frac{\lambda_c}{\lambda_s} \ll 1$. This is significant as it provides an explicit  perturbative solution to the strongly-interacting collective dynamics of the Kuramoto system in its synchronized states, beyond the asymptotic (and thermodynamic) limit $\lambda \to \infty$.

\section{Concluding remarks} \la{conclusions}

In future work, we will explore the universality class limit of this model, around the planar solution (at equilibria and otherwise) and around the special solution $\vec{S}_j = \widehat{e}_3$. Having an embedding of the original model into a family of Hamiltonian dynamical systems allows to explore non-equilibrium features of this dynamical phase transition in the formalism of integrability, which is in itself a nontrivial connection between theories usually seen as not naturally compatible. 

%\section*{Acknowledgments}

\end{document}